\documentstyle[prl,aps,twocolumn]{revtex}

\begin{document}
\twocolumn[\hsize\textwidth\columnwidth\hsize\csname @twocolumnfalse\endcsname

\title{
Competition between spin and charge polarized states \\
in nanographene ribbons with zigzag edges
}
\author{
Atsushi~Yamashiro,$^1$ 
Yukihiro~Shimoi,$^{1,2}$
Kikuo~Harigaya,$^{1,2,*}$
and Katsunori~Wakabayashi$^{3}$
}
\address{
$^1$Nanotechnology Research Institute, AIST, 1-1-1 Umezono, Tsukuba 305-8568, Japan \\
$^2$Research Consortium for Synthetic Nano-function Materials Project, AIST, 1-1-1 Umezono, Tsukuba 305-8568, Japan \\
$^3$Department of Quantum Matter Science, Graduate School of Advanced Sciences of Matter, Hiroshima University,  Higashi-Hiroshima 739-8530, Japan
}
\date{\today}

\maketitle

\begin{abstract}
\ Effects of the nearest neighbor Coulomb interaction on nanographene ribbons with zigzag edges are investigated using the extended Hubbard model within the unrestricted Hartree-Fock approximation. The nearest Coulomb interaction stabilizes a novel electronic state with the opposite electric charges separated and localized along both edges, resulting in a finite electric dipole moment pointing from one edge to the other. 
This charge-polarized state competes with the peculiar spin-polarized state caused by the on-site Coulomb interaction and is stabilized by an external electric field.\\

\noindent DOI:10.1103/PhysRevB   ~~~~~~~~~~~~~~~~~~~~~~~~~~PACS number(s):71.10.Hf, 73.22.-f,73.20.At,77.22.Ej
\end{abstract}

\vskip2pc] 

\narrowtext  
New materials composed of carbon atoms have been attracting much attention both in the fundamental science and in the potential application to nanotechnology devices\cite{Iijima91,DresselhausDE,SaitoDD98}. One of the main interests is the variety of their physical and chemical natures depending on geometries
\cite{Iijima91,DresselhausDE,SaitoDD98,FujitaWNK96,HikiHLM,WakaSF,WakaHari03,Harigaya01,HarigayaE01}. 
In carbon nanotubes,  diameters and chiral arrangements of hexagonal pattern on tubules decide whether they are metallic or not\cite{Iijima91,DresselhausDE,SaitoDD98}.
Besides the closed carbon molecules like carbon nanotubes, carbon systems with open boundaries also show unusual features.

In nanographenes, i.e., nanoscale graphites, their peripheral edges strongly affect the electronic states\cite{FujitaWNK96}: A nanographene ribbon with zigzag edges (zigzag ribbon) has the nonbonding molecular orbitals localized mainly along the zigzag edges (edge states). They constitute partly flat bands and cause a sharp peak in the density of states at the Fermi energy. On the other hand, one with armchair edges does not have such edge states. When the on-site Coulomb interaction is introduced within the unrestricted Hartree-Fock (UHF) approximation, it causes the Fermi-instability of the flat bands and forms the spin-polarized (SP) states: most spin density concentrates along the zigzag edges forming the ferrimagnetic orders due to the edge states. As to the effects of the Coulomb interactions, there still remains the qualitatively important problem of the competition between the spin and charge orderings in this system due to the on-site and nearest neighbor Coulomb interactions, respectively. 
In fact, for two dimensional (2D) boundless graphite, the possibility of the stable charge density wave (CDW) state compared with the spin density wave (SDW) state was reported\cite{WagnerL,TchougreeffH}.

In this paper, we demonstrate that the nearest neighbor Coulomb interaction stabilizes a novel charge-polarized (CP) state with a finite electric dipole moment in zigzag ribbons  and it competes with the SP state.

Figure 1 illustrates the geometry of a zigzag ribbon with an inversion symmetry.
$N$ and $L$ are the width and length of the ribbon, respectively. 
The dashed rectangle denotes a unit cell. The periodic boundary condition is set along the $y$ axis. 
Since the zigzag ribbon is bipartite, A and B sites are assigned by black and white circles, respectively. Note that all the twofold coordinated sites in the lower and upper zigzag edges belong to the A  and B sublattices regardless of $N$, respectively, at which edge states are mainly localized. 

  We treat a half-filled  $\pi$ electron system on the zigzag ribbon  with the extended Hubbard Hamiltonian,
\begin{eqnarray}
H_{EH}  & = &  -t\sum_{\langle i, j \rangle, s}(c_{i,s}^{\dagger}c_{j,s} + {\rm H.c.})\nonumber\\
       & + &  U\sum_{i}(n_{i,\uparrow} -\frac{1}{2})(n_{i,\downarrow} -\frac{1}{2})
     +  V\sum_{\langle i, j \rangle}(n_{i}- 1)(n_{j} - 1).
\end{eqnarray}
 Here $\langle i,j\rangle$ denotes a pair of neighboring carbon sites and $-t$  is the transfer integral between them. 
 The operator $c^{\dagger}_{i,s}$ creates a $\pi$ electron with spin $s$ on $i$-th site and $c_{i,s}$ annihilates the one. 
 $n_{i,s}$ is the corresponding number operator and $n_{i}=\sum_{s}n_{i,s}$. The parameters $U$ and $V$ denote the on-site and nearest neighbor Coulomb interactions, respectively. 

We adopt the standard UHF approximation\cite{FujitaWNK96} and the mean field Hamiltonian is
\begin{eqnarray}
h=&-&\sum_{\langle i, j\rangle,s}[(t+V\langle c_{j,s}^{\dagger}c_{i,s}\rangle)c_{i,s}^{\dagger}c_{j,s}+{\rm H.c.} - V|{\langle}c_{i,s}^{\dagger}c_{j,s}\rangle|^{2} ]\nonumber\\
&+&U\sum_{i}(\sum_{s}{\langle}n_{i,-s}-\frac{1}{2}{\rangle}n_{i,s}-{\langle}n_{i,\uparrow}{\rangle}{\langle}n_{i,\downarrow}\rangle+\frac{1}{4}\;)\nonumber \\
&+&V\sum_{{\langle}i, j{\rangle}}({\langle}n_{i}-1{\rangle}n_{j}+{\langle}n_{j}-1{\rangle}n_{i}-{\langle}n_{i}{\rangle}{\langle}n_{j}\rangle+1).
\end{eqnarray}
The charge density at $i$-th site is given by $d_{i}=1-\langle n_{i}\rangle$  while  the $z$ component of spin density at $i$-th site is given by $s_{i}=(\langle{n}_{i,\uparrow}\rangle-\langle n_{i,\downarrow}\rangle)/2.$

Figure 2 (a) shows the charge density distribution in the CP state with $N{\times}L=4{\times}20$, where $U=0.3t$ and $V=0.4t$. For comparison, Fig. 2 (b) shows the spin density profile of the SP state for $U=t$ and $V=0.$ The CP state has no spin density at every site, while the SP state has no charge density at every site. 
In the CP state,  the lower zigzag edge are charged positively, while the upper zigzag edge are charged negatively.
The distribution pattern of the charge density in the CP state is just like that of the spin density in the SP state. The charge polarization in the CP state is explained by the interplay between the edge states and the Fermi-instability of the flat bands due to $V$ as follows, just as the spin polarization explained by the edge states and $U$\cite{FujitaWNK96}.
First, $V$ stabilize the charge ordered state similar to the CDW state with A and B sites charged positively and negatively, respectively, that is the Fermi-instability of the flat bands due to $V.$
Therefore, all the twofold sites in the lower zigzag edge, A sites, are charged positively, while those in the upper zigzag edge, B sites, are charged negatively. Moreover, magnitudes of the charge densities at twofold sites are much larger than those at adjacent threefold sites with unlike charges due to the edge states. 
Therefore, both zigzag edges are charged oppositely and the CP state is regarded as a peculiar CDW state caused by the edge states.

The CP state has a finite electric dipole moment pointing from one edge to the other along the $x$-axis. 
It is formed in the homogeneous system and breaks the inversion symmetry. 
The $x$ component of polarization operator is 
$P_{x} = ea\sum_{i}{\bar{x}_{i}}(1-n_{i}),$ 
where $e$ and $a$ are the elementary electric charge and the C-C bond length in a zigzag ribbon, respectively. 
$\bar{x}_{i}$ is the $x$ coordinate of the $i$-th site in unit of $a$.  The electric dipole moment per unit cell $\mu = \langle{P}_{x}\rangle/(L/2)$ in Fig. 2 (a) is 0.88 [$ea$] (= 5.9 Debye for $a = 1.4\; {\rm \AA}$). 

 Figure 3  shows the phase diagram in the parameter space of $U$ and $V$ for the competition between the CP and SP states, where $N{\times}L=4{\times}40$.
Above the phase boundary the CP state is lower in energy than the SP state, otherwise the SP state is lower.
The boundary is determined as the crossing points of their energy curves by the numerical interpolation. 
Near the boundary, the CP and SP states coexist for  given parameter sets, showing the first order phase transition. 

At $U{\sim}0$, the boundary rises up with the infinite gradient, indicating that the CP state is higher in energy than the SP state in the weak coupling limit. Since Fig. 2 suggests that the magnitudes of charge densities become vanishingly small at threefold sites while they remain finite at twofold sites,  the CP state cannot be stabilized by $V$ and is simply destabilized by $U$. This comes from the edge states absent in nanographene ribbons with armchair edges and in 2D boundless graphites.%\cite{TchougreeffH}.

When $V$  exceeds $V_{c}{\sim}t/2$ below the phase boundary, the CP state gradually loses their localized nature and the magnitudes of charge densities grow continuously at threefold sites. In the strong coupling limit, the magnitude of the charge density becomes almost the same among all sites and the CP state turns into the CDW state, showing the crossover from the CP state to the CDW state.
This is just like the crossover from the SP state to the SDW state at $U_{c}{\sim}t$ above the phase boundary\cite{FujitaWNK96}.
In the CDW state, the magnitudes of unlike charge densities are the same so that the total charge along each edge cancels out, respectively, while ${\langle}P_{x}{\rangle}$ remains finite due to the non-uniform width of the zigzag ribbons. 
In the continuous change to the CDW state, ${\langle}P_{x}{\rangle}$ increases to approach $LN/4$.

The inset of Fig. 3 shows the phase diagram in a reduced scale. 
In the strong coupling limit, the phase boundary between the CDW and SDW states approaches the line $U={\eta}V$ derived from equalizing their Coulomb energies per site $E_{CDW}=U/4-V{\eta}/2$ and $E_{SDW}=-U/4,$ where $\eta=3-1/N$ is the average number of adjacent sites which originates from threefold sites in the unit cell with $2N$ sites except for {\it two} twofold sites at edges. Note that the CDW-SDW phase boundary in 1D or 2D system is generally close to the line with $\eta$ being the number of adjacent sites for each system\cite{TchougreeffH,ZhangC,Hirsch,Nakamura,SenguptaSC,TsuchiizuF,Jeckelmann}, while the edge states in a zigzag ribbon cause a new feature, i.e., the deviation of the phase boundary from it in the weak coupling limit. In the wide ribbon limit, the phase boundary approaches $U=3V$ for 2D graphite since the weights of the edges states diminish and $\eta$ approaches 3.

Finally,  we demonstrate how an external electric field stabilizes the CP state.
Since the external field couples with $\mu$, the CP state with finite electric dipole moment $\mu_{CP}$ is effectively stabilized while the SP state with vanishing electric dipole moment $\mu_{SP}$ not.
Under the static electric field with the amplitude $E_{ext}$ polarized along the $x$ axis, Hamiltonian is given by
$H = H_{EH} - E_{ext}P_{x}.$
The unit of $E_{ext}$ is $t/(ea)$ which is $7.1{\times}10^{7} \;{\rm V/cm}$ for $t=1\;{\rm eV}$.
Figure 4 (a) shows the total energies of the CP and SP states as functions of $E_{ext}$ for $U=0.9t$ and $V=0.6t$. 
At $E_{ext}=0$, the SP state is lower in energy than the CP state.
Applying the electric field, the SP's energy depends parabolically on $E_{ext}$ with the peak at the origin and scarcely decreases. This is because the initial vanishing $\mu_{SP}$ eliminates the first order perturbation energy. Concomitantly, the induced $\mu_{SP}$ is proportional to $E_{ext}$ as shown in Fig. 4 (b). On the other hand, the CP's energy depends linearly on $E_{ext}$ and decreases appreciably, because the initial $\mu_{CP}$ is much larger than the induced one and the first order energy correction is dominant. Note that the CP state becomes lower in energy than the SP state when $E_{ext}{\geq}4.5{\times}10^{-3} \;[t/(ea)].$ Figure 4 (b) shows discontinuous change of $\mu$  at the transition, due to the first order one.

The followings are devoted to discussions.
First, the nanographene ribbon is regarded as a quasi 1D system where quantum fluctuations 
are important and an UHF solution is not a reliable ground state.
Such a system generally has several UHF solutions with almost degenerate energies and 
the natures of their long range orders appear as short range orders in the correlated ground state\cite{IkawaF}.
On the other hand, the parameter set $U=2.5t$ and $V=t$ for 2D graphite borrowed from the coronen molecule\cite{TchougreeffH} falls into the SP side in the vicinity of the phase boundary in Fig. 3.
This suggests that the CP and SP states in zigzag ribbons are close in energy and 
there will be the charge and spin fluctuations along zigzag edges reflecting the CP and SP states.
In fact, Hikihara {\it et al}\cite{HikiHLM} showed that the correlated ground state in the Hubbard model for zigzag ribbons is the spin-singlet state without symmetry breaking while the ferrimagnetic correlation at the zigzag edges remains robustly.
We expect that the CP state is a stable UHF solution as the SP state is proved to be in the random phase approximation\cite{WakaSF}.
The symmetry breaking $\mu_{CP}$ is due to the UHF approximation and the inversion symmetry will be restored in the correlated ground state with $\mu=0$. It will include the degenerate CP states with oppositely signed $\mu_{CP}$ as important configurations and exhibit large quantum fluctuation of $\mu$ similar to the quantum paraelectricity\cite{Barrett,Horiuchi}. 

Secondly, the electron correlation will change the nature of the phase transition obtained in the UHF approximation.
In a 2D extended Hubbard model with square lattice, the quantum Monte Carlo study showed that the transition is similar to that in 1D system\cite{ZhangC}, where extensive studies clarified that in the weak and strong coupling limits there are continuous (second order) and discontinuous (first order) phase transitions at $U\simeq2V$, respectively\cite{Hirsch,Nakamura,SenguptaSC,TsuchiizuF,Jeckelmann}.
This may hold for the quasi 1D zigzag ribbon, while the divergent gradient of the boundary at $U{\sim}0$ will survive, i.e., the charge fluctuations reflecting the CP states will be suppressed by the edge states in the weak coupling limit under the correlation since the edge states are robust against the correlation\cite{WakaSF,HikiHLM}.

Thirdly, this model is based on  
the screened Coulomb interactions that effectively take into account the correlation beyond the UHF approximation as seen in studies of polyacetylene\cite{Fukutome}. As to the effect of the long range Coulomb interactions, the {\it ab initio} study in 2D graphite suggests that it prefers the CDW state\cite{WagnerL}.  
However, it may destabilize the CP state since the long range Coulomb repulsions between like charges along the same edges prevail those attractions between unlike charges along the opposite edges. If the excess repulsive energy surpasses the energy gain due to the charge ordering in this model, the CP state will be destabilized. 
Even if it is the case, the CP state is stabilized under the electric field as shown in Fig. 4 and will affect the dielectric response of the system.
The effects of the long range Coulomb interactions and the correlation will be studied in future. 

Fourthly, the effect of the geometrical topology on the electronic state is another interesting problem.
Recently, Tanda {\it et al.} synthesized M{\"o}bius strips formed by the single crystals of $\rm NbSe_{3}$\cite{Tanda}. 
This work urged theoretical studies on the effects of the M{\"o}bius boundary condition\cite{WakaHari03,YakuboAC,HayashiE}.
In the zigzag ribbon, this topology causes a ferrimagnetic domain wall between the up and down spin domains localized along zigzag edges within the Hubbard-UHF model\cite{WakaHari03}.  
$V$ induces another novel state in a M{\"o}bius strip: the ferroelectric domain wall connecting the domains with the opposite electric dipole moments\cite{YamashiroYHW}. 
The competition between these states in the exotic topology will be studied elsewhere.

Finally, the present results are not only applied to the nanographene ribbons but also to boron-nitride nano-ribbons\cite{Chen,HarigayaBN}.
We conjecture that the novel CP state is valid for the general 2D $\pi$ electron systems with zigzag edges and  affects their dielectric response to be different from those without zigzag edges.

In summary, we found the novel charge-polarized state with a finite electric dipole moment caused by the interplay between the nearest neighbor Coulomb interaction and the edge states in nanographene ribbons with zigzag edges using the extended-Hubbard-UHF model.
This state competes with the spin-polarized state caused by the on-site Coulomb interaction and the former is effectively stabilized by an external electric field.

This work has been supported partly by Special Coordination Funds for Promoting Science and Technology and by NEDO under the Nanotechnology Materials Program.
\begin{figure}[ht]
\begin{center}
%\epsfile{file=fig1.ps} 
\caption{
Schematic structure of a bipartite nanographene ribbon with zigzag edges (zigzag ribbon); $\bullet$, A site; $\circ$, B site. Dashed rectangle denotes a unit cell. 
}
\end{center}
\end{figure} 
\begin{figure}[ht]
\begin{center}
%\epsfile{file=fig2.ps} 
\caption{
(a) Charge density distribution, $d_{i}$, of the charge-polarized (CP) state for $U=0.3t$ and $V=0.4t,$ and (b) the $z$ component of spin density distribution, $s_{i}$, of the spin-polarized (SP) state for $U=t$ and $V=0,$ on a zigzag ribbon with $4{\times}20$ sites; $\bullet$, positive density; $\circ$, negative density.
Radii of circles denote the magnitudes of the charge or spin densities; the maximal one is 0.18 in (a) and 0.12 in (b).
}
\end{center}
\end{figure} 
\begin{figure}[ht]
\begin{center}
%\epsfile{file=fig3.ps}
\caption{
Phase diagram in the parameter space of $U$ and $V$ for a zigzag ribbon composed of $4{\times}40$ carbon atoms. Solid curve is the phase boundary between the CP and SP states which fits numerical data ($\bullet$). The inset shows the phase diagram in a reduced scale. The dashed line denotes the CDW-SDW phase boundary in the strong coupling limit. 
}
\end{center}
\end{figure}  
\begin{figure}[ht]
\begin{center}
%\epsfile{file=fig4.ps}
\caption{
External static electric field $E_{ext}$ dependence of (a) energies, and (b) electric dipole moment $\mu$ of 
the CP state ($\bullet$) and SP state ($\circ$). 
}
\end{center}
\end{figure}

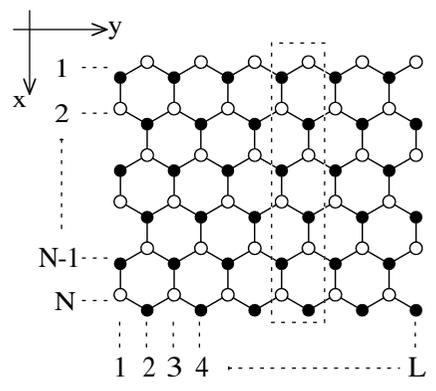
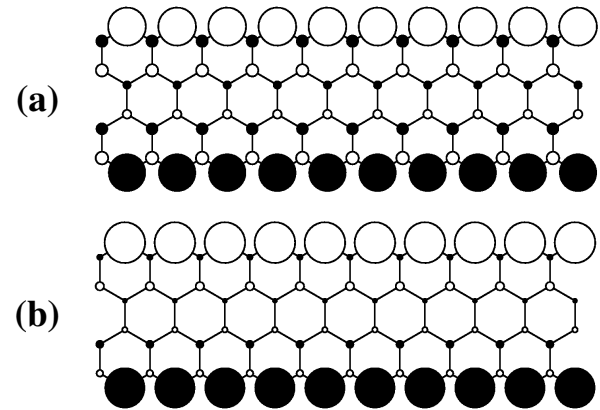
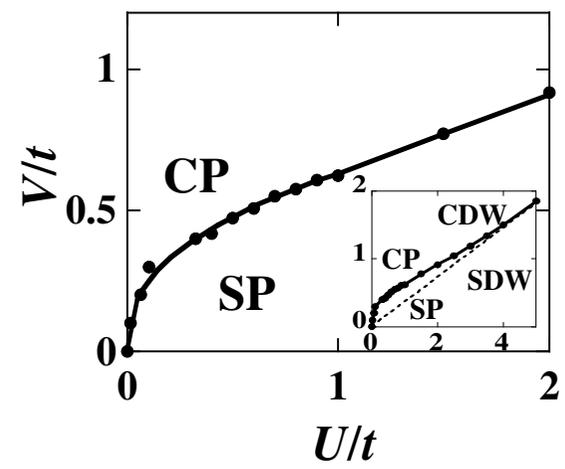
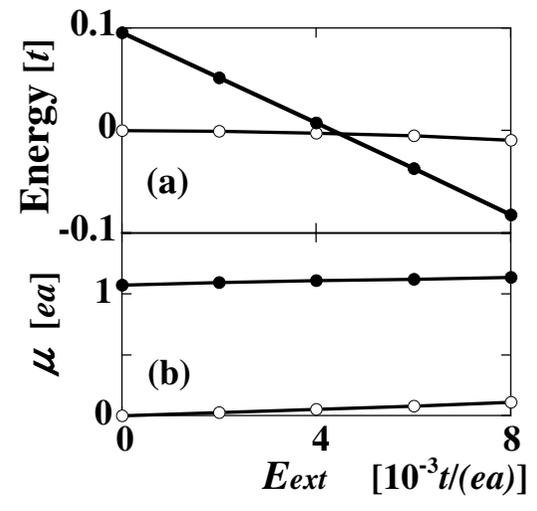
\end{document}